\documentclass[aip,jcp, floatfix, reprint]{revtex4-1}

\usepackage{amsmath}
\usepackage{graphicx}
\usepackage[T1]{fontenc}
\usepackage{bm}
\usepackage{dcolumn}
\usepackage{color}
\usepackage{lipsum}
\usepackage{gensymb}
\usepackage{hyperref}
\usepackage{etoolbox}

\apptocmd{\thebibliography}{\raggedright}{}{}

\def\maxfloatwidth{%
  \ifdim\columnwidth>246.0pt
  300.0pt  \else
  \columnwidth
  \fi
}

\newcommand{\subs}[1]{$_{#1}$}

\newcommand{\trm}[1]{\textrm{#1}}

\begin{document}

\title{The Microscopic Features of Heterogeneous Ice Nucleation May
  Affect the Macroscopic Morphology of Atmospheric Ice Crystals}

\author{Stephen J. Cox} 
\affiliation{Thomas Young Centre and Department of Chemistry,
  University College London, London WC1E 6BT, U.K.}
\affiliation{London Centre for Nanotechnology, University College
  London, 17-19 Gordon Street, London WC1H 0AH}

\author{Zamaan Raza} 
\affiliation{Thomas Young Centre and Department of Chemistry,
  University College London, London WC1E 6BT, U.K.}

\author{Shawn M. Kathmann} 
\affiliation{Physical Sciences Division, Pacific
  Northwest National Laboratory, Richland, Washington 99352, United
  States}

\author{Ben Slater} 
\affiliation{Thomas Young Centre and Department of Chemistry,
  University College London, London WC1E 6BT, U.K.}

\author{Angelos Michaelides} 
\affiliation{Thomas Young Centre and Department of Chemistry,
  University College London, London WC1E 6BT, U.K.}
\affiliation{London Centre for Nanotechnology, University College
  London, 17-19 Gordon Street, London WC1H 0AH}
\email{angelos.michaelides@ucl.ac.uk}

\date{\today}

\begin{abstract}

It is surprisingly difficult to freeze water. Almost all ice that
forms under ``mild'' conditions (temperatures > \(-40\)\celsius)
requires the presence of a nucleating agent - a solid particle that
facilitates the freezing process - such as clay mineral dust, soot or
bacteria. In a computer simulation, the presence of such ice
nucleating agents does not necessarily alleviate the difficulties
associated with forming ice on accessible timescales. Nevertheless, in
this work we present results from molecular dynamics simulations in
which we systematically compare homogeneous and heterogeneous ice
nucleation, using the atmospherically important clay mineral kaolinite
as our model ice nucleating agent. From our simulations, we do indeed
find that kaolinite is an excellent ice nucleating agent but that
contrary to conventional thought, non-basal faces of ice can nucleate
at the basal face of kaolinite. We see that in the liquid phase, the
kaolinite surface has a drastic effect on the density profile of
water, with water forming a dense, tightly bound first contact
layer. Monitoring the time evolution of the water density reveals that
changes away from the interface may play an important role in the
nucleation mechanism. The findings from this work suggest that
heterogeneous ice nucleating agents may not only enhance the ice
nucleation rate, but also alter the macroscopic structure of the ice
crystals that form.

\end{abstract}

\maketitle

\section{Introduction}
\label{sec:intro}

Ice formation is a process important to numerous fields, ranging from
microbiology\cite{kim:afp-pnas, protein-mimicry, Christner29022008} to
understanding and predicting chemical processes in the atmosphere
where, for example, it is known that ice particles in the polar
stratospheric regions catalyse the formation of radicals responsible
for ozone depletion.\cite{abbatt:atmos-review} Almost all ice
formation is facilitated by the presence of a (solid) foreign body, in
a process known as \emph{heterogeneous nucleation} and it is well
reported that different materials affect the rate of ice formation to
different extents.\cite{PK97, cantrell2005, zimmermann2008,
  eastwood2008, murray-review} However, despite the wide ranging
consequences of ice formation, little is understood about how the
surface properties of a foreign body affect its ice nucleating
ability. By furthering our knowledge of the microscopic details of
heterogeneous nucleation, it is possible that new pathways to the
rational design of materials that either inhibit or enhance ice
formation can be explored, with implications for the
atmospheric\cite{demott:1997, demott:1998, murray:cirrus,
  baker:science, verlinde:2007, mcfarquhar:2007, murray:nature} and
climate sciences,\cite{choi:2006, lee:jatmossci} along with the food
and transport industries.

Whereas experiments aimed at measuring the ice nucleating ability of
different materials relevant to the atmosphere provide useful
information for global climate models, as well as telling us which
materials actually make good ice nucleating agents, most of our
molecular level understanding of water--surface interactions are a
result of detailed surface science studies (for an overview see
Refs.~\onlinecite{ThielMadey, henderson-review, hodgson-review,
  salmeron-review, ewing-review, javi-review,
  angelos:faraday2007}). For example, through the combined use of
density functional theory (DFT) calculations and experiments such as
scanning tunnelling microscopy and infrared spectroscopy, the
structures of the first wetting layer at
Pt(111)\cite{feibelman:2010:prl, feibelman:2010:jcp} and sub-monolayer
chains at Cu(110)\cite{javi:pentagons} have been elucidated. Such
studies are, however, unable to provide the simultaneous spatial and
temporal resolution required to probe the heterogeneous ice nucleation
mechanism, making computer simulation an appropriate tool to study
such a process. Despite a number of computer simulation studies of
homogeneous nucleation,\cite{matsumoto:nature, jungwirth:subsurf,
  jungwirth:surfactant, jungwirth:salt, molinero:isitcubic,
  molinero:nomans, doye:umbrella, trout:umbrella, quigley:154518}
there have been very few that have directly probed heterogeneous
nucleation. Yan and Patey\cite{patey2011, patey2012} have performed an
excellent set of molecular dynamics (MD) simulations aimed at
investigating the effect of strong electric fields on ice nucleation,
finding that ferroelectric cubic ice forms in the region exposed to
the electric field. Although this provides some insight into the role
of electric fields on nucleation, the fields used are relatively
smooth, whereas those exerted by real surfaces are likely to greatly
vary on molecular length scales. Solveyra \emph{et
  al.}\cite{molinero:confinement} have also looked at the effect of
confinement on ice nucleation in both hydrophilic and hydrophobic
nanopores, using the single-site mW water
model.\cite{molinero:mW-orig} Use of such coarse grained force fields
to describe the molecular interactions has the distinct advantage of
being able to simulate large length- and time-scales at reasonable
computational cost, but would unfortunately be inappropriate for the
current study, where the electrostatic interactions between the
surface and water are significant. The work presented here is unique
in that, to our knowledge, it will be the first to directly simulate
the dynamical process of heterogeneous nucleation where the atomic
structure of both water and the substrate is taken into account.

As with homogeneous nucleation, there are many computational
techniques at our disposal for looking at heterogeneous
nucleation. One possible route is to use a free-energy based method
such as metadynamics\cite{orig-metadyn} or umbrella
sampling\cite{Torrie1977187} (for applications of these methods to
homogeneous ice nucleation see Refs.~\onlinecite{quigley:154518,
  trout:umbrella, doye:umbrella}). The advantage of methods such as
these is that one is able to obtain free energy barriers to nucleation
along a specified reaction coordinate, but with the drawback that the
system has to be driven along a predetermined set of collective
variables, with no guarantee that the `true' reaction pathway is being
sampled. Another approach is to perform a number of unbiased MD
simulations, starting with water in the supercooled liquid state, over
suitably long time-scales until the nucleation event is
observed. Although adopting such an approach may be seen as
computationally inefficient, with recent advances in computer
technology and software, the timescales involved are realisable at a
reasonable computational cost for small to medium system
sizes. Furthermore, by only performing unbiased MD simulations, we are
no longer imposing \emph{a priori} the reaction coordinate that the
system must traverse. This direct approach has been used to seemingly
good effect to study homogeneous ice nucleation, first by Matsumoto
\emph{et al.}\cite{matsumoto:nature} and subsequently by Jungwirth and
co-workers.\cite{jungwirth:subsurf, jungwirth:surfactant,
  jungwirth:salt}

With our aim of understanding heterogeneous ice nucleation, we have
opted to explore the clay mineral kaolinite as our model ice
nucleating agent. Each year, as much as 3000~Tg of mineral dust
(naturally occurring crystalline solid compounds) is transported into
the troposphere from desert regions\cite{ipcc2001} where it catalyses
the formation of ice.\cite{PK97,murray-review} The composition of
mineral dust is diverse with quartz, feldspar, calcite and clays all
present in significant proportions in typical atmospheric dust
samples. Clays are the most frequently observed group in atmospheric
mineral dust, of which kaolinite forms a substantial
fraction.\cite{murray-review} Apart from being a known effective ice
nucleating agent,\cite{eastwood2008, zimmermann2008, murray:kaolinite}
the binding of water to the pristine hydroxyl-terminated (001) face
has been well characterised theoretically,\cite{xiaoliang2007,
  xiaoliang2008} which aids in the analysis of our nucleation
simulations.

\begin{figure}[htb]
  \centering
  \includegraphics[width=0.8\linewidth]{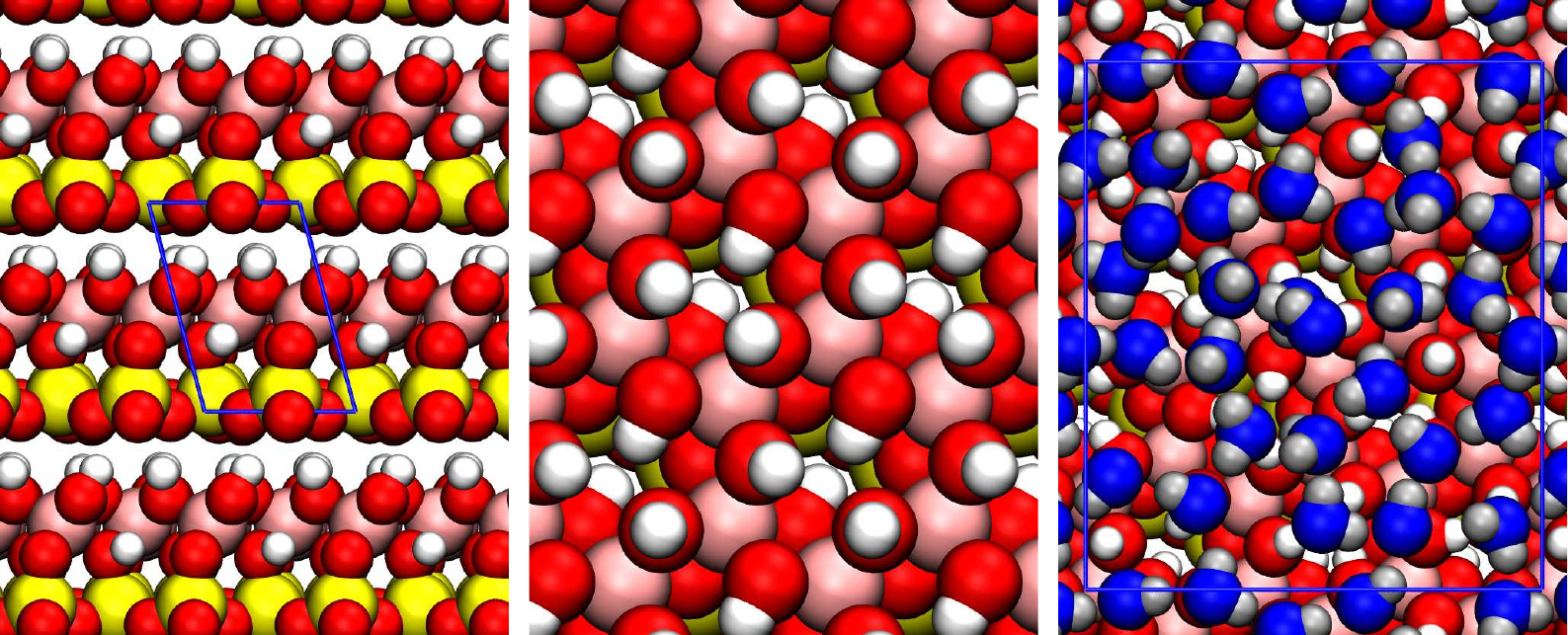}
  \caption{\textbf{Structure of kaolinite.} On the left we show the
    layered bulk structure of kaolinite. As the layers are bound by
    hydrogen bonds between the hydroxyl-terminated and
    silicate-terminated faces, facile cleavage is observed along in
    the (001) plane. The middle panel shows the hydroxyl-terminated
    (001) face. DFT calculations\cite{xiaoliang2007, xiaoliang2008}
    show that upon cleavage, 1/3 of OH groups rotate into the plane of
    the surface, making it amphoteric i.e. able to both accept and
    donate hydrogen bonds with water. On the right is a snapshot from
    one of the MD simulations showing the first contact layer of
    supercooled water. We can see that the water molecules are densely
    packed and disordered. The colour scheme is: Si, yellow; Al, pink;
    O, red; and H, white. Water molecules in the first contact layer
    are shown in blue.}
  \label{fig:kao}
\end{figure}

Kaolinite is a layered silicate mineral with chemical composition
Al\(_{2}\)Si\(_{2}\)O\(_{5}\)(OH)\(_{4}\). Each layer consists of a
tetrahedral silica sheet alternating with an octahedral alumina sheet,
terminated with hydroxyl groups (see Fig.~\ref{fig:kao}). In the bulk,
these layers are bound by hydrogen bonds between the
hydroxyl-terminated face and the silica-terminated face, giving rise
to facile cleavage along the (001) plane, exposing the hydroxyl- and
silicate-terminated faces. It is believed that the hydrophilic
hydroxyl-terminated face is the origin of ice nucleating efficacy of
kaolinite, with the textbook explanation being that the
pseudo-hexagonal arrangement of --OH groups acts as a template upon
which the basal face of ice I\subs{\trm{h}} can grow.\cite{PK97}
Despite its attractive simplicity, the validity of this explanation
has been questioned; a series of DFT calculations by Hu and
Michaelides\cite{xiaoliang2007, xiaoliang2008, xiaoliang2010} indicate
that the most stable ice-like bilayer at the kaolinite surface is
actually hydrophobic with respect to growth of further layers of ice,
a property attributed to the amphoteric nature of the
hydroxyl-terminated surface; whilst grand canonical Monte Carlo
simulations by Croteau \emph{et al.}\cite{patey2008, patey2009,
  patey2010} have shown that only small regions of hexagonal motifs
form in the first water overlayer and that these are somewhat
stretched relative to bulk ice. In this work, we will directly probe
the ice nucleation mechanism at the kaolinite (001) surface using MD
simulations in a bid to shed further light onto the process of ice
formation in the presence of this important mineral, as well as make
inroads into understanding heterogeneous ice nucleation in general.

In what follows, we will see that, rather than the basal face, we
exclusively see the prism face growing from the kaolinite surface. We
will show that density fluctuations in the supercooled water away from
the kaolinite slab play an important role in the heterogeneous
nucleation mechanism. We will also discuss the role finite size
effects play in our simulations before concluding and discussing the
implications for the macroscopic crystal structure of our findings.

\section{Methods}
\label{sec:methods}

As we wish to simulate many molecules over long time scales it will be
necessary to use classical force fields as opposed to quantum
mechanical methods such as DFT. To this end, we employ the TIP4P/2005
water model\cite{vega:tip4p-2005} and the CLAYFF potential of Cygan
\emph{et al.}\cite{cygan:clayff} to describe the
kaolinite. TIP4P/2005, a rigid point charge water model, has been
shown to replicate the phase diagram of water qualitatively well along
with the transport properties of bulk water, even though it predicts
the melting point of ice I\subs{\trm{h}} to be \(ca.\) 252~K. It also
reproduces the experimental bulk densities of liquid water, hexagonal
and cubic ice very well, making it a suitable choice for modelling ice
nucleation. The CLAYFF potential has been widely used for studying
water at various clay mineral interfaces\cite{cygan2007, cygan2005,
  acs:clayff, chandler:talc} and in particular for the study of ice
nucleation at kaolinite by grand canonical Monte
Carlo.\cite{patey2008, patey2009, patey2010} In this approach, the
clay atoms are treated as simple point charges with Lennard-Jones
interactions, with the only explicit bonding term occurring between
the oxygen and hydrogen of the hydroxyl groups. Such flexibility in
the model allows CLAYFF to describe a number of different clay
structures and phases satisfactorily, as well as the swelling of clays
with increased water content.\cite{cygan:clayff} The water-clay
interaction was calculated using the standard Lorentz-Berthelot mixing
rules.\cite{lorentz-mix, berthelot-mix}

For the MD simulations, we followed a protocol similar to that used by
Jungwirth and co-workers,\cite{jungwirth:subsurf,
  jungwirth:surfactant, jungwirth:salt} who have had much success in
direct simulation of homogeneous ice nucleation. To create our
homogeneous systems, 192 water molecules were placed in an orthogonal
simulation cell with lateral (\(xy\)) dimensions of \emph{ca.} \(13.2
\times 15.6\)~\AA\(^{2}\).\footnote{The lateral cell dimensions for
  the homogeneous simulations were obtained from a 0.01~K NPT
  simulation of a proton ordered configuration of hexagonal ice. In
  the case of the heterogeneous simulations, the lateral cell
  dimensions are constrained to be commensurate with the kaolinite
  slab. The kaolinite structure used was based upon the experimental
  structure of Bish\cite{bish:1993}, with the \(\alpha\) and
  \(\gamma\) angles altered slightly to make the cell orthogonal.} Due
to the small \(x\)- and \(y\)-dimensions, a small cutoff of 6.5~\AA\,
was employed. Electrostatic interactions were calculated using the
smooth particle mesh Ewald method, with a pseudo 2D
correction\cite{yeh:2dEwald} for the slab geometry, giving an
effective \(z\)-dimension of at least 100~\AA. The geometry of this
system can thus be best described as an infinite slab with two
liquid-vapour interfaces. For the heterogeneous system, the kaolinite
was modelled as a single slab and 192 water molecules were placed on
the hydroxyl-terminated (001) face, creating a solid-liquid interface,
whilst leaving a liquid-vapour interface. Due to the presence of the
kaolinite substrate, the lateral dimensions were \emph{ca.} \(15.5
\times 17.9\)~\AA\(^{2}\), slightly larger than in the homogeneous
case. To ensure that the kaolinite slab did not drift, one of the
silicon atoms was fixed throughout the simulations.

To propagate the dynamics, the velocity Verlet algorithm was used with
a timestep of 2~fs. Simulations were performed in the canonical
ensemble and the temperature was controlled using a Nos\'{e}-Hoover
chain of length 10 and a temperature coupling constant of 0.5~ps. Both
systems were equilibrated at 300~K for 2~ns from which initial
configurations for the production runs were sampled. For the
production simulations, the systems were quenched to 220~K
(i.e. approximately 30~K supercooled) and ran for the order of
1~\(\mu\)s or until nucleation was observed. The water geometry was
maintained using the SETTLES\cite{settles} algorithm, whereas the
P-LINCS\cite{hess:lincs} algorithm was used to constrain the O--H bond
in kaolinite. All simulations were performed using the {\scriptsize
  GROMACS 4.5} simulation package.\cite{gromacs4}

\section{Results and Discussion}
\label{sec:results}

\begin{figure}[htb]
  \centering
  \includegraphics[width=0.8\linewidth]{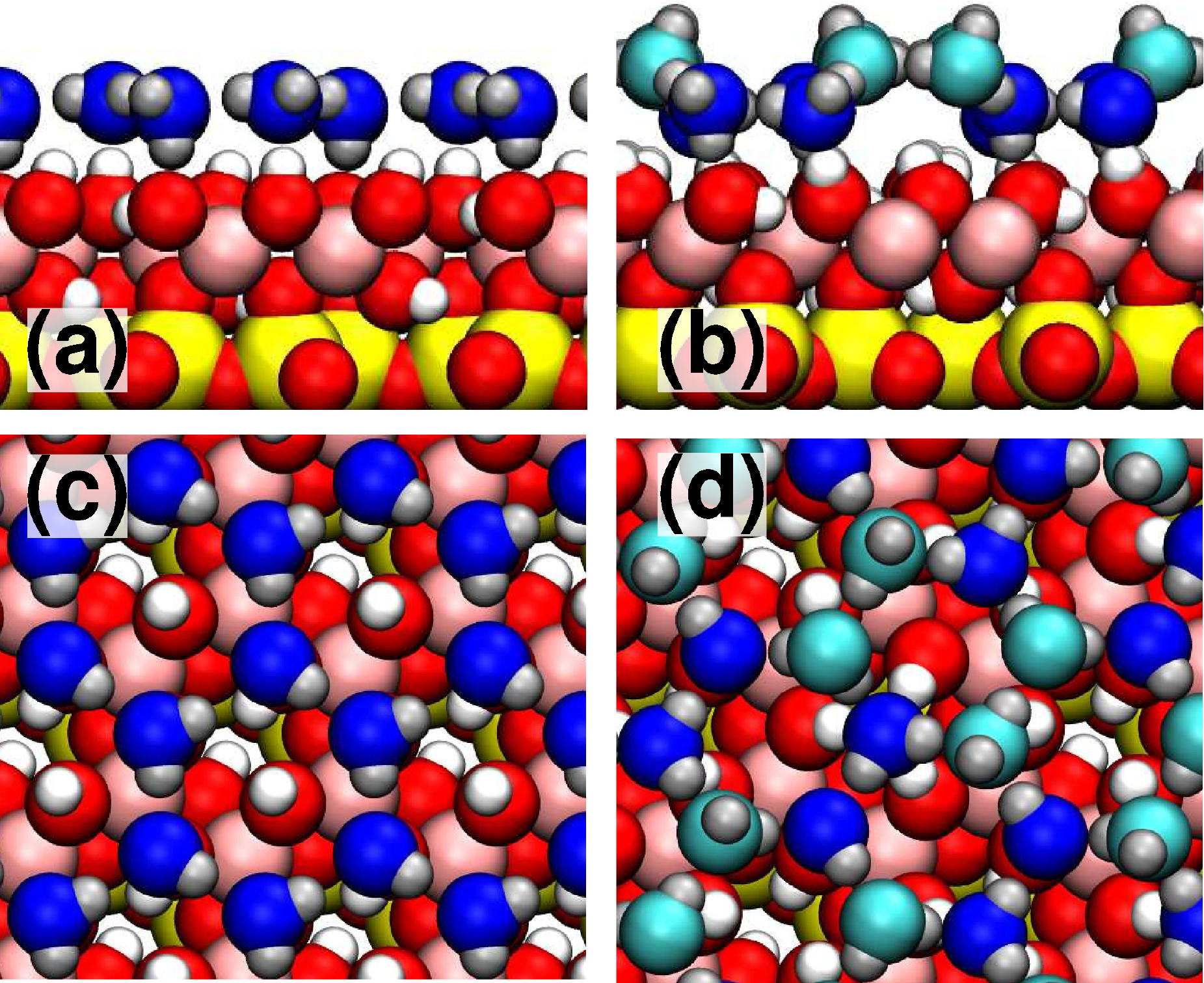}
  \caption{\textbf{Diagram of ice-like structures at the kaolinite
      surface.} In panel (a) we show a side view of the basal face of
    ice bound to kaolinite in the ``H-down bilayer''
    configuration.\cite{xiaoliang2008} All water molecules bind with
    similar heights from the surface. Panel (b) shows a side view of
    the prism face bound to kaolinite. In this structure, the water
    molecules come in high-lying (light blue) and low-lying (dark
    blue) pairs. Note that the prism face structure donates hydrogen
    bonds to the surface, as well as having `dangling' hydrogen bonds
    pointing away from the surface (these dangling hydrogen bonds are
    absent in the basal face structure). Panels (c) and (d) show top
    views of the basal and prism face structures, respectively.}
  \label{fig:layers}
\end{figure}

In total we ran 27 heterogeneous simulations, observing 10 nucleation
events and 30 homogeneous simulations, observing 9 nucleation
events. Movies of some of these are available in the Supporting
Information. Before doing any detailed analysis one trend was
immediately clear: on the kaolinite we exclusively formed hexagonal
ice whereas in the homogeneous simulations we generally observed a
mixture of hexagonal and cubic stacking patterns. In all but one of
the homogeneous simulations, at least half of the ice formed consisted
of cubic sequences with the bilayers parallel to the liquid/vapour
boundary. Only one simulation resulted in solely hexagonal ice. This
is qualitatively consistent with X-ray diffraction data and Monte
Carlo simulations performed by Malkin \emph{et al.}, which
demonstrated that the homogeneous nucleating phase is stacking
disordered (denoted as ice I\subs{\trm{sd}}), consisting of roughly
equal numbers of cubic and hexagonal
sequences.\cite{murray:cubic-pnas, ice_sd:pnas} This mixture of cubic
and hexagonal layers is also consistent with previous simulation
studies.\cite{jungwirth:subsurf, carignano:stacking} Furthermore, when
ice forms homogeneously we see a variety of crystal orientations
within the simulation cell whereas when ice forms on the kaolinite, we
always see growth along the prism face of ice and not the basal face
i.e. the ice bilayers grow perpendicular to the kaolinite slab. In
Fig.~\ref{fig:layers} we show diagrams of the basal and prism faces at
the kaolinite surface. The observation that the prism face nucleates
at kaolinite is interesting, as it means that the pseudo-hexagonal
arrangement of --OH groups at the kaolinite surface are not acting as
a template for the basal face of ice.

\begin{figure*}[htb]
  \centering
  \includegraphics[width=0.75\linewidth]{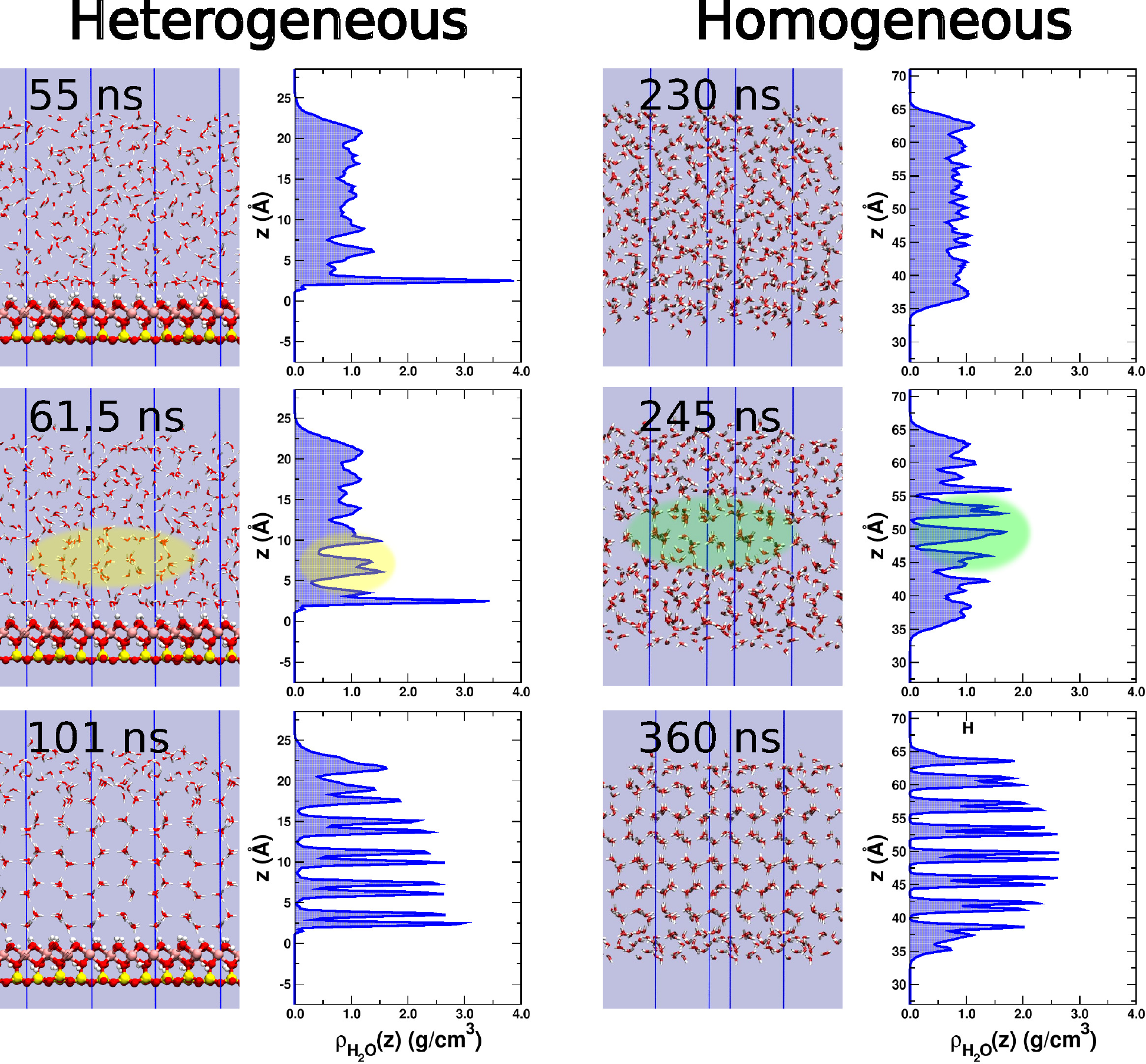}
  \caption{\textbf{Snapshots and water density profiles for a
      homogeneous (right) and heterogeneous (left) nucleation event.}
    In the presence of kaolinite, the supercooled water (55~ns) has a
    high density peak corresponding to the first contact layer. There
    is also a noticeable second peak, but this is far less intense and
    much broader. After 61.5~ns, nucleation has started. There is a
    slight reduction in density of the first density peak, but this is
    still much higher than anywhere else in the system. Rearrangement
    of water molecules in the second layer associated with a split in
    the density peak (highlighted in yellow), is also seen and is
    indicative of an ice-like layer forming. By 101~ns the first
    contact layer has fully transformed to ice and the density is
    similar to that observed in the rest of the system. Note that it
    is the prism face of ice exposed to the kaolinite surface and that
    we only observe hexagonal ice. In contrast, for the homogeneous
    slab we see a fairly uniform density profile in the supercooled
    regime (230~ns). We also see a mixture of hexagonal and cubic
    stacking. In this instance, the initial nucleation event
    (highlighted in green) leads to a cubic stacking arrangement. The
    densities are averages over a 2.5~ns interval centred at the
    specified time. The colour scheme is the same as
    Fig.~\ref{fig:kao}.}
  \label{fig:main-fig}
\end{figure*}

From visual inspection of the ice-forming trajectories it was noticed
that, during the nucleation event, considerable rearrangement of the
water molecules always seemed to occur in the second water layer above
the kaolinite surface (note that this statement does not preclude any
rearrangement occurring in the first or third layers). To provide
evidence for this observation we measured how the density of water
varies with height along the \(z\)-direction during the transition. In
Figure~\ref{fig:main-fig} we present this analysis for a single
heterogeneous and homogeneous simulation along with the corresponding
snapshots. First of all, we can see that the supercooled liquid (shown
at 55~ns) has an extremely sharp and intense density peak at the
kaolinite surface, as well as a pronounced, but broader, second
peak.\footnote{This high density peak is also a feature of water at
  kaolinite at 300~K, which we show in the Supporting Information,
  along with a density profile from a DFT-MD simulation at 330~K that
  also exhibits such a peak.} After 61.5~ns the nucleation event has
occurred and we can see the intensity of the first peak has decreased
slightly, although it still remains much higher than anywhere else in
the system. We also see that the second peak has started to split
(highlighted in yellow), indicative of an ice-like layer forming. It
is only after this change in density in the second layer that we see
the first layer transform fully to ice. We can compare this to the
homogeneous case, where the density in the supercooled liquid is
essentially uniform and the nucleation event seems to occur by two or
three layers concurrently forming ice. In both the homogeneous and
heterogeneous scenarios, once the initial nucleation event has
occurred the growth of ice then proceeds, with a quasi liquid-like
layer remaining at the water/vapour interface, consistent with
previous simulation studies.\cite{jungwirth:subsurf,
  carignano:stacking, ding:acsnano, slater:faraday} We note that the
observed changes away from the surface have striking similarity with
the previously reported `collective mechanism' for ice growth along
non-basal faces at temperatures below 240~K.\cite{Nada-collective,
  kusalik:ice-growth:2012}

\begin{figure*}[htb]
  \centering
  \includegraphics[width=0.8\linewidth]{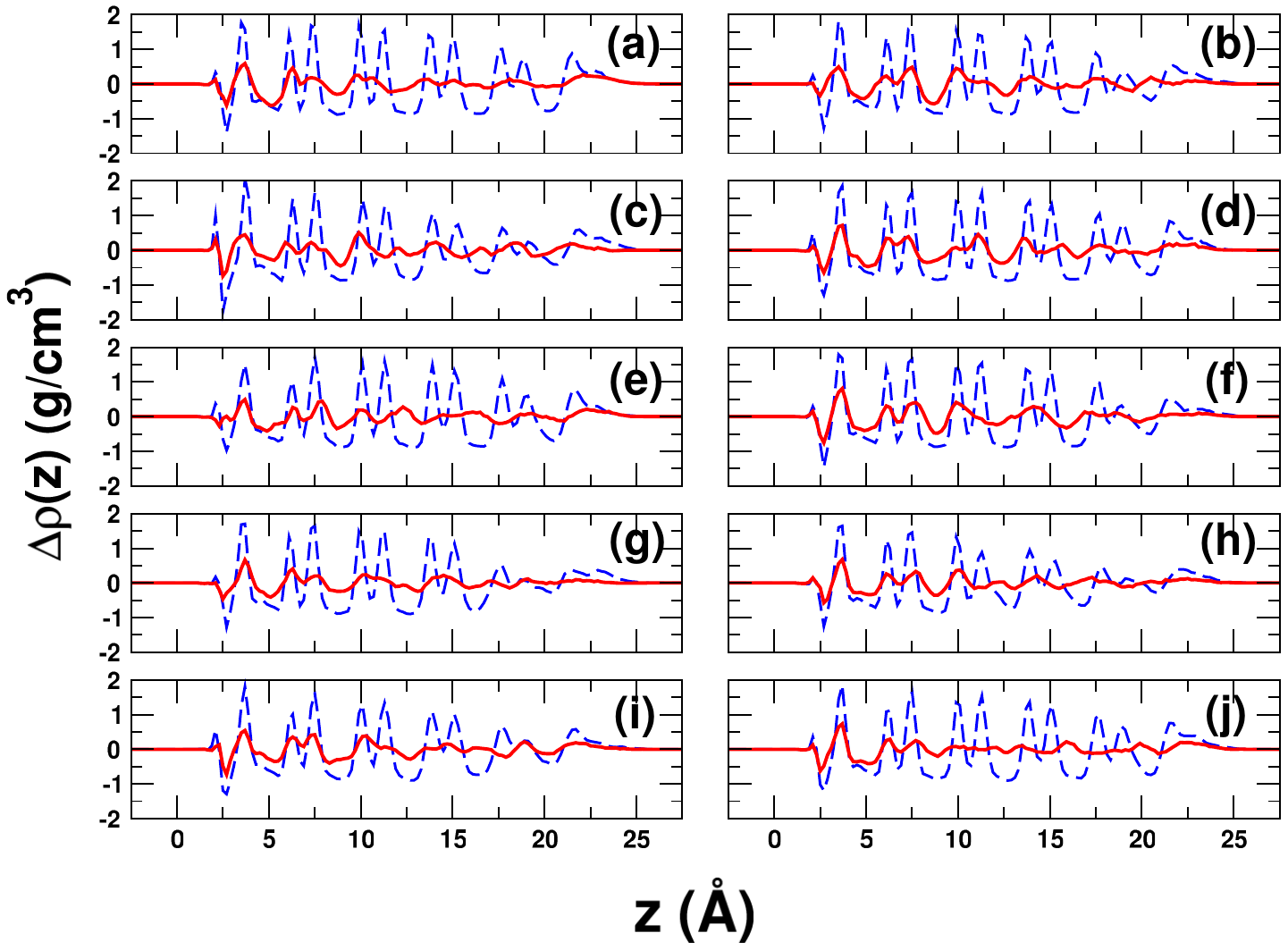}
  \caption{\textbf{Water density difference profiles for all
      heterogeneous nucleation events.} Each panel (a-j) shows an
    independent nucleation event. The quantity plotted is
    \(\Delta\rho(z)\) as defined by Equation~\ref{eqn:dendiff}. The
    red solid line shows \(\Delta\rho(z)\) at a time just after the
    onset of nucleation and the blue dashed line shows
    \(\Delta\rho(z)\) at a later time after ice has grown. In all
    cases, we see that there are density changes in the second layer
    (just below 7.5~\AA\,) of a similar size to those in the first
    layer, before ice goes on to form fully (noticeable changes in the
    third layer are also often observed). In the case of (b), we know
    from a committor analysis that the red line corresponds to a
    pre-critical configuration. The displayed densities are averages
    over 2.5~ns.}
  \label{fig:all-den}
\end{figure*}

To investigate these structural changes away from the surface further,
for each heterogeneous nucleation event observed we have computed the
density difference:

\begin{equation}
  \label{eqn:dendiff}
  \Delta\rho(z) = \rho(z) - \langle\rho_{\trm{liq}}(z)\rangle
\end{equation}

\noindent where \(\rho(z)\) is the instantaneous water density at a
height \(z\) and \(\langle\rho_{\trm{liq}}(z)\rangle\) is the water
density at a height \(z\) averaged over supercooled liquid
configurations. The results are presented in Fig.~\ref{fig:all-den}
(for reference, panel (b) corresponds to the heterogeneous nucleation
event presented in Fig.~\ref{fig:main-fig}). We can clearly see that
in all instances, just after the onset of nucleation, there is a
change in the density of the second (and often the third) water layer
that is comparable to the changes observed in the first layer. In none
of the simulations do we observe the first layer fully transform to
ice without this signature splitting of the second layer density.

\begin{figure}[htb]
  \centering
  \includegraphics[width=0.8\linewidth]{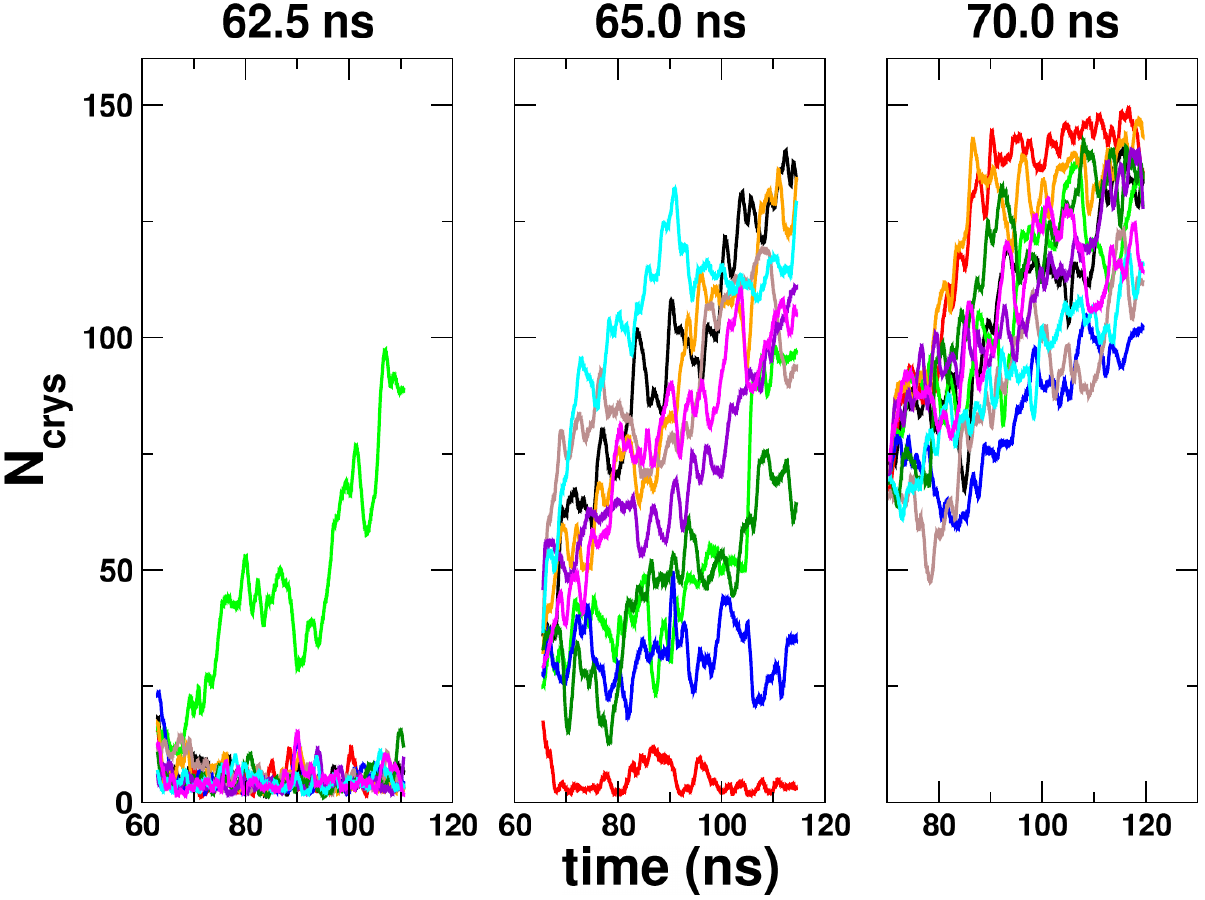}
  \caption{\textbf{Committor analysis from one of the heterogeneous
      ice nucleation trajectories.} Results here are shown for initial
    configurations sampled at 62.5~ns, 65.0~ns and 70.0~ns from the
    initial ice forming trajectory. 10 independent trajectories were
    started from each configuration by giving the particles random
    velocities sampled from a Maxwell-Boltzmann distribution. By
    monitoring the number of water molecules defined as being ice by
    the CHILL algorithm,\cite{molinero:chill} \(N_{\trm{crys}}\), we
    are able to determine whether or not ice forms. We can clearly see
    that at 62.5~ns we are in a pre-critical regime and by 70~ns all
    trajectories continue to form ice. At 65.0~ns we do not see all
    trajectories form ice or liquid, but some stay somewhere in
    between the two states, even over the \emph{ca.} 50~ns
    timescale. Results are presented as running averages over a 1~ns
    interval.}
  \label{fig:committor}
\end{figure}

It is important to consider how significant the changes in density
away from the surface are in the ice nucleation mechanism; after all,
one may argue that these are just a consequence of the initial changes
seen in the first layer and are merely indicative of ice growth rather
than playing a role in the nucleation mechanism itself. We have
therefore performed a committor analysis on the heterogeneous
trajectory presented in Fig.~\ref{fig:main-fig} (and panel (b) in
Fig.~\ref{fig:all-den}), using the CHILL algorithm of Moore \emph{et
  al.}\cite{molinero:chill} to monitor ice formation. This was done by
choosing different configurations along this trajectory and starting
10 new trajectories with random velocities drawn from the
Maxwell-Boltzmann distribution. Results from three starting
configurations are presented in Fig.~\ref{fig:committor}, where we can
clearly define a pre- and post-critical region (initial configurations
from 62.5~ns and 70.0~ns of the initial trajectory, respectively). In
between these two regimes, however, we do not see an expected 50:50
split of trajectories going on to reach the liquid and ice states,
rather we see some that definitely go to ice, some that definitely go
to liquid but some trajectories that stay somewhere in between, even
over fairly long timescales (\emph{ca.} 50~ns). As the cost of this
committor analysis is high, we have not attempted to refine our search
further and remain satisfied that the configuration sampled at 65.0~ns
is a reasonable representation of the `transition region'. What is
relevant to our discussion regarding the density changes in the second
layer is that \(\Delta\rho(z)\) shown by the red line shown in
Fig.~\ref{fig:all-den}(b) corresponds to the configuration sampled at
62.5~ns i.e. the splitting in the second peak for this trajectory
occurs in the pre-critical regime, indicating that these structural
changes are part of the nucleation mechanism rather than a feature of
growth. We have also performed a similar analysis for the trajectory
in Fig.~\ref{fig:all-den}(d), which we present in the Supporting
Information, along with movies showing how \(\Delta\rho(z)\) varies
during the nucleation event.

\begin{figure}[htb]
  \centering
  \includegraphics[width=0.8\linewidth]{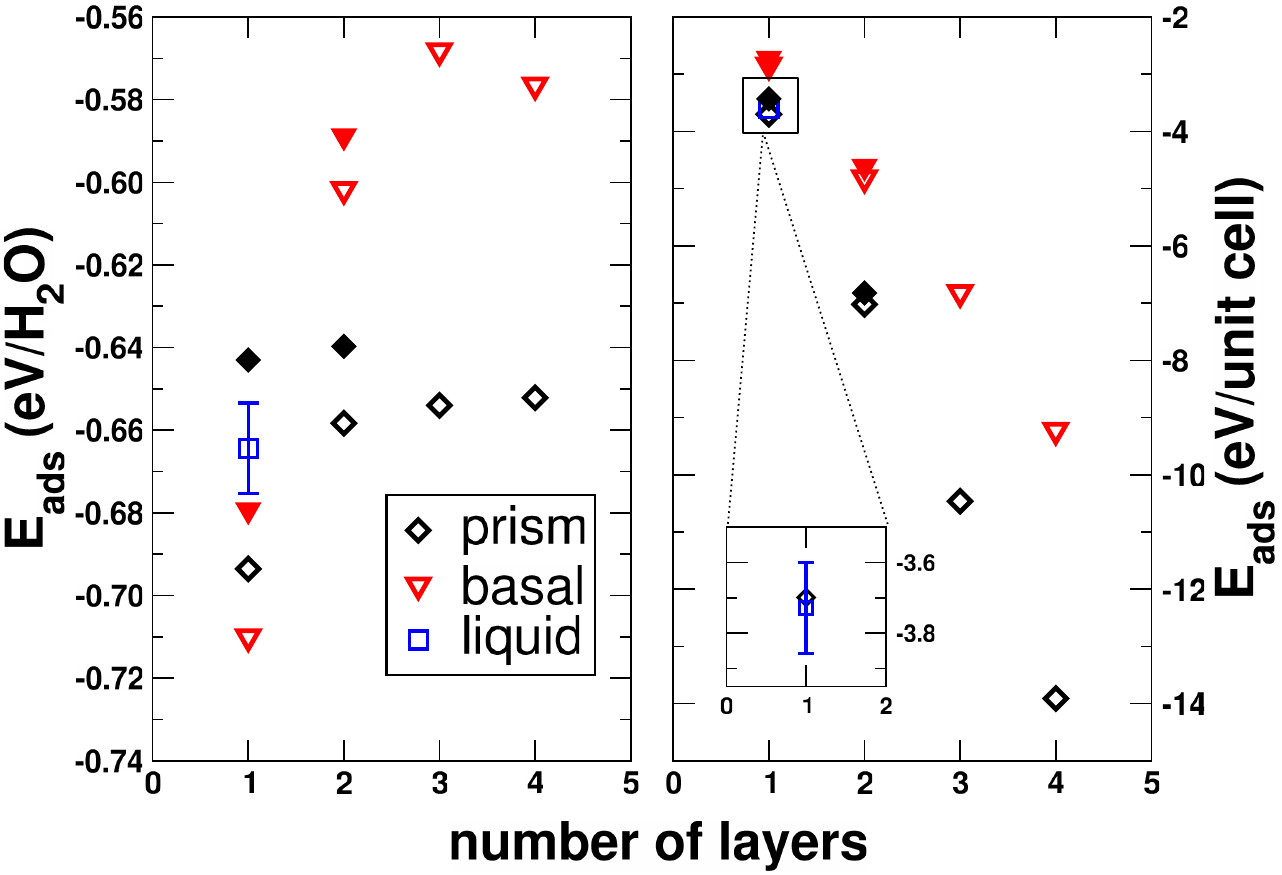}
  \caption{\textbf{Variation of the adsorption energy
      (\(E_{\trm{ads}}\)) of ice to kaolinite as the number of ice
      layers changes.} The black diamonds show results for ice binding
    to kaolinite through its prism face, whilst the red triangles show
    results for ice binding through its basal face. Filled symbols
    show results from DFT calculations. The left panel shows the
    adsorption energy calculated per water molecule, whereas the right
    panel shows the adsorption energy per conventional unit cell of
    kaolinite. For the first contact layer on its own, the adsorption
    energy per water molecule is stronger for the basal face than the
    prism face, but upon adsorption of other layers, the prism face
    structure becomes significantly more stable. When ice binds
    through the prism face, the coverage of water molecules is higher
    than when it binds through the basal face, meaning that the
    adsorption energy per unit cell of kaolinite is more stable for
    the prism face independent of the number of adsorbed layers. Data
    for the first liquid layer is also shown (the bars indicate
    estimates of the thermal fluctuations). On a per molecule basis,
    this is less stable than the ice-like structures, but is more
    stable per unit cell of kaolinite.}
  \label{fig:ener-vs-layer}
\end{figure}

It is interesting to attempt to explain some of these observations. To
help understand why we see the formation of ice with its prism rather
than basal face exposed to the kaolinite surface, we have investigated
how the adsorption energy of ice changes with the number of ice-like
layers, for both the prism and basal faces bound to the kaolinite
(details of these calculations are given in the Supporting
Information). The results of this analysis are presented in
Fig.~\ref{fig:ener-vs-layer}, where we present the data in terms of
adsorption energy per water molecule and adsorption energy per
conventional unit cell of kaolinite. When only the first contact layer
is present, the basal face of ice is more strongly bound than the
prism face by approximately 15~meV/H\(_{2}\)O. As soon as we go beyond
the first layer, however, the prism face becomes more stable, with the
difference becoming more pronounced as more layers are added. The
prism face also binds with a higher coverage than the basal face (5.33
vs. 4 H\(_{2}\)O per conventional unit cell) meaning that the prism
face is more stable per unit cell of kaolinite independent of the
number of ice-like layers. To understand these differences, it is
useful to examine the structure of the ice-like layers when binding
through the prism and basal faces, which we show in
Fig.~\ref{fig:layers}. Here it can be seen that the water molecules in
the basal face structure bind with similar heights from the kaolinite,
with half the molecules donating one hydrogen bond to the surface and
the other half accepting a hydrogen bond from the kaolinite whilst
donating two hydrogen bonds to other water molecules (the ``H-down
bilayer'' structure as described in
Ref.~\onlinecite{xiaoliang2008}).\footnote{The ``H-up bilayer''
  structure was also tested, but this was always less stable than the
  ``H-down'' structure and so has been omitted for clarity.}  By
adopting this structure, the water molecules maximise their bonding to
the kaolinite and maintain good hydrogen bonding between each other,
giving a large overall adsorption energy for the first layer. This
structure, however, saturates all hydrogen bonds leaving no `dangling'
hydrogen bonds that water molecules in above layers can bind to and
consequently, the adsorption energy rapidly becomes less exothermic as
other layers are added. This finding is consistent with previous
findings from a DFT study,\cite{xiaoliang2008} as well as the
experimental observation that the availability of dangling hydrogen
bonds determines the multilayer wetting behaviour of water on metal
and metal oxide surfaces.\cite{salmeron:faraday} On the other hand,
the prism face binds with a somewhat more corrugated configuration,
with the water molecules coming in high-lying and low-lying pairs. One
of the molecules in the low-lying pair donates one hydrogen bond to
the kaolinite whilst its partner accepts hydrogen bonds from the
kaolinite. The high-lying pairs bridge the low-lying pairs through
hydrogen bonds, with the important feature that one of these
high-lying molecules has an OH bond directed away from the surface
i.e. the prism face exhibits dangling hydrogen bonds. The fact that
half of the molecules come in high-lying pairs means that the
adsorption energy per water molecule of the first layer is less for
the prism face than it is for the basal face, but the ability of the
prism face to donate and accept hydrogen bonds to both the surface and
the above water layers means that it becomes more stable as the number
of water layers increases. We have also computed the adsorption
energies of the first and second layers with DFT using the
Perdew-Burke-Ernzerhof (PBE) exchange-correlation
functional\cite{PBE-GGA} (full details of these calculations are given
in the Supporting Information). Although agreement is not exact
between our force field setup and PBE (which should not be taken as a
benchmark) the trend that prism face becomes more stable than the
H-down bilayer upon adsorption of a second layer of ice is still
seen. This suggests that this observation is not an artifact of our
choice of force field.

In Fig.~\ref{fig:ener-vs-layer} we also show the average adsorption
energy of the first layer from 25 configurations selected from the
supercooled liquid.\footnote{Water molecules are defined as being in
  the first layer if their oxygen atoms are within 5~\AA\, of the
  average height of the kaolinite surface oxygen atoms. This
  corresponds to a cutoff that is approximately halfway between the
  first and second ice layers.} On a per molecule basis, the liquid
layer is less stable than either of the ice-like structures, but from
the right hand panel of Fig.~\ref{fig:ener-vs-layer} we can see that
per unit cell of kaolinite, the liquid layer is slightly more
stable. This result may help us explain the observed density changes
away from the surface during the nucleation process. If we draw an
analogy to the grand canonical ensemble, we may consider the first
water layer as a subsystem that is able exchange heat and particles
with the bulk liquid above.\footnote{We must emphasise that this is an
  analogy to the grand canonical ensemble and that the first water
  layer and the above liquid are strongly coupled.} In the supercooled
state, therefore, there will be some pseudo-equilibrium number of
water molecules in the first layer, which we have measured to be 5.61
H\(_{2}\)O/unit cell (c.f. 5.33 H\(_{2}\)O/unit cell for the prism
face). Thus, although the adsorption energy per water molecule is
stronger for the first layer of ice, on average more water molecules
are present in the first liquid layer leading to an overall
stabilisation. For ice to form and persist at the surface, it is
therefore required that the average number of water molecules at the
surface decreases. In keeping with the analogy to the grand canonical
ensemble, this amounts to a need for a change in the chemical
potential of the reservoir of water molecules above the first layer,
which manifests itself as the structural changes away from the surface
discussed previously. We can also see from the right panel of
Fig.~\ref{fig:ener-vs-layer} that the adsorption energy of the prism
face per unit cell is within our estimate of the thermal fluctuations
from the average liquid value. This may be one of the reasons for
kaolinite's good ice nucleating ability.

Finally, it is important that we mention the role of finite size
effects in this work. We have attempted to perform these simulations
with system sizes doubled in the lateral dimensions (768 water
molecules, cutoff for interactions extended to 9~\AA) but no
nucleation was observed in a total simulation length of 15.5~\(\mu\)s
over a temperature range spanning of 190--220~K. We also simulated
2~\(\mu\)s at 240~K using the TIP4P/ice water
model,\cite{abascal:234511} which has a melting point similar to
experiment, but still no nucleation was observed. This discrepancy can
be explained by the fact that in the small systems, there is a self
interaction of the growing ice nucleus with its periodic images that
lowers the interfacial free energy cost of nucleation. We have also
performed 14 homogeneous simulations in the same cell used for the
heterogeneous simulations (i.e. lateral dimensions of \emph{ca.}
\(15.5 \times 17.9\)~\AA\, without the kaolinite slab). No nucleation
events were observed. Although it would have been desirable to have
observed nucleation in these simulations, so that we could have
compared homogeneous and heterogeneous rates, one pleasing aspect of
this last null result is that it means that the heterogeneous
nucleation results presented earlier are not completely dominated by
finite size effects. As we are able to routinely observe nucleation in
this cell when the kaolinite slab is present, but not homogeneously,
we are left to conclude that kaolinite significantly enhances the
rate. We are not currently in a position, however, to go beyond this
qualitative level.

As a final test of the finite size effects, we doubled the lateral
cell dimensions and used a configuration from one of our heterogeneous
simulations, replicated in both dimensions to fill the larger cell, as
an initial configuration. Taking the configuration that we have
determined to be representative of the transition region from the
committor analysis as a `seed' configuration for the larger cell, we
still see ice growth in the same manner as the small cells. The fact
that we see growth and not a collapse of the crystal suggests that the
prism face is stable on the hydroxyl-terminated (001) kaolinite face
and is not solely stabilised by periodic boundary effects present in
the small cells.

\section{Conclusions}
\label{sec:conlcu}

We have investigated ice nucleation in thin water films, both
homogeneously and heterogeneously in the presence of a kaolinite slab,
using regular molecular dynamics simulation. We have performed many
simulations on the order of one microsecond, observing many nucleation
events. In agreement with previous simulation studies and recent
experiments, in the case of homogeneous nucleation we see a mixture of
cubic and hexagonal arrangements. Contrary to expectation, at the
kaolinite surface we always see growth along the prism face of ice,
suggesting that the source of kaolinite's good ice nucleating ability
does not lie with its good epitaxial match with the basal face of
ice. By monitoring the density of water above the kaolinite slab
during the nucleation event, we see that changes in the second water
layer appear crucial to the nucleation mechanism. The growth of the
prism face rather than the basal face is due to the ability of the
former to bind favourably to both the surface and water layers above,
as well as having a higher coverage. The observed structural changes
away from the surface have been explained as allowing the average
number of water molecules in the first layer to decrease, which
subsequently allows the remaining water molecules to form the favoured
ice-like structure. We have, however, seen that finite size effects
are non-negligible in these simulations, with no nucleation observed
upon moving to bigger cells. Nevertheless, the fact that we do not
observe homogeneous nucleation in the cell size used for the
heterogeneous nucleation simulations suggests that the results on
kaolinite are not entirely dominated by the finite size effects. This
result also shows that kaolinite is a potent ice nucleating agent.

Given the finite size effects, it would be highly desirable to
implement a free energy method that could definitively probe the
heterogeneous nucleation mechanism proposed here. Even with the
current state-of-the-art in free energy methods and advanced sampling
techniques, freezing water is still likely to be difficult. The reason
for this is that slow dynamics offered by the hydrogen bonding network
present in supercooled water makes it very difficult for methods such
as umbrella sampling and metadynamics to equilibrate the system as it
is pushed along the chosen order parameter.\cite{doye:all-atom}
Furthermore, advanced sampling techniques that exploit natural
dynamics, such as transition path sampling\cite{TPS:big-review} or
forward flux sampling\cite{FFS:review} are likely to suffer as the
actual transition time is relatively long (tens of nanoseconds, as
seen in Fig.~\ref{fig:committor}), which may make sampling
computationally prohibitive. One way to circumvent this problem is
through the use of a coarse grained potential such as the mW
model\cite{molinero:mW-orig} which, by treating hydrogen bonding in a
mean-field sense, reduces the complexity of the underlying potential
energy surface and results in faster dynamics. This has already been
used to good effect with both direct molecular dynamics (see
e.g. Ref.~\onlinecite{molinero:nature}) and forward flux
sampling\cite{galli:mWnuc} for homogeneous nucleation. Such methods
could be used to verify previous homogeneous simulations that suffer
from similar finite size effects.\cite{jungwirth:subsurf,
  jungwirth:surfactant, jungwirth:salt} This approach is unlikely to
work in the case of heterogeneous nucleation on substrates such as
clays, however, where electrostatics are dominant. How to proceed in
such cases is at present unclear, but given the industrial and
environmental implications of ice formation, the topic deserves a
major research effort.

Finally, the fact that the pristine kaolinite surface promotes the
growth of the prism face over the basal face may have consequences for
the macroscopic crystal structure of ice that forms. Ice exhibits a
complex habit diagram\cite{hallett:ice-habit} and as the surface
cleavage energies of the prism and basal faces are very
similar\cite{ding:cleavage} it is possible that different
heterogeneous ice nucleating agents could tip the balance to favour
different ice habits under the same conditions. As the macroscopic
structure of an ice crystal can affect its light scattering
properties, understanding the effect of ice nucleating agents may be
important for global climate models. Future calculations will probe
the influence of other ice nucleating agents on nucleation and growth
processes with the aspiration of comparing with measurements from
cloud chamber experiments.

\section{Acknowledgements}
\label{sec:ack}

The authors would like to thank Dr. Xiao-Liang Hu for many helpful
discussions when commencing this work and for providing Fig. S2. This
research used resources of the National Energy Research Scientific
Computing Center, which is supported by the Office of Science of the
U.S. Department of Energy under Contract No. DE-AC02-05CH11231. We are
grateful to the London Centre for Nanotechnology and UCL Research
Computing for computational resources.  Via our membership of the UK's
HPC Materials Chemistry Consortium, which is funded by EPSRC
(EP/F067496), this work made use of the facilities of HECToR, the UK's
national high-performance computing service, which is provided by UoE
HPCx Ltd. at the University of Edinburgh, Cray Inc., and NAG Ltd., and
funded by EPSRC's High End Computing Programme. The authors also
acknowledge the use of the UCL Legion High Performance Computing
Facility, and associated support services, in the completion of this
work. S.M.K. was supported fully by the U.S. Department of Energy,
Office of Basic Energy Sciences (BES), Division of Chemical Sciences,
Geosciences, and Biosciences. S.J.C. was supported by a student
fellowship funded jointly by UCL and BES. A.M. is supported by the
European Research Council and the Royal Society through a Royal
Society Wolfson Research Merit Award.

\bibliography{model-hex}

\begin{thebibliography}{87}
\bibitem [{Noted()}]{Noted}%
  \BibitemOpen
  \bibinfo {note} {Water molecules are defined as being in the first layer if
  their oxygen atoms are within 5~\r A\protect \tmspace +\thinmuskip {.1667em}
  of the average height of the kaolinite surface oxygen atoms. This corresponds
  to a cutoff that is approximately halfway between the first and second ice
  layers.}\BibitemShut {Stop}%
\bibitem [{Notee()}]{Notee}%
  \BibitemOpen
  \bibinfo {note} {We must emphasise that this is an analogy to the grand
  canonical ensemble and that the first water layer and the above liquid are
  strongly coupled.}\BibitemShut {Stop}%
\end{thebibliography}

\end{document}